\documentclass{article}

\usepackage{arxiv}

\usepackage[utf8]{inputenc} 
\usepackage[T1]{fontenc}    
\usepackage{hyperref}       
\usepackage{url}            
\usepackage{booktabs}       
\usepackage{amsfonts}       
\usepackage{nicefrac}       
\usepackage{microtype}      
\usepackage{lipsum}
\usepackage{graphicx}
\usepackage{amsmath}
\usepackage{xcolor}

\title{Fundamental diagram of vibration-driven vehicles}

\author{
 German A. Patterson \\
  Instituto Tecnológico de Buenos Aires (ITBA), CONICET\\
  C.A. de Buenos Aires, Argentina \\
  \texttt{gpatters@itba.edu.ar} \\
   \And
 Daniel R. Parisi \\
  Instituto Tecnológico de Buenos Aires (ITBA), CONICET\\
  C.A. de Buenos Aires, Argentina \\
  \texttt{dparisi@itba.edu.ar} \\
}

\begin{document}
\maketitle
\begin{abstract}

In this study, we conducted experimental investigations into the fundamental diagram of vibration-driven vehicles (VDV) in a one-dimensional array. As these mechanical agents interact solely through collisions, their mean speed remains nearly constant at low and medium densities. However, there is a reduction of between 25\% and 40\% when the lineal density approaches the inverse of the contact distance. Remarkably, in this one-dimensional system, the outcome is significantly influenced by the order in which agents, sorted by their free speeds, are gradually introduced into the experiment. While a significant speed difference is observed at low and medium densities based on this ordering, both curves eventually converge to the same speed at maximum density. Moreover, the attained speed in saturated systems is slower than the speed of the slowest agent.

\end{abstract}


\section{Introduction}

The fundamental diagram for a set of self-propelled particles relates either the flow rate or the mean velocity of these particles to their density. It has been extensively used for characterizing vehicle traffic systems and as a tool for their design \cite{newell1961nonlinear, pipes1966car, schadschneider1997traffic, schadschneider2000statistical, siebel2006fundamental, helbing2009derivation, yao2019stability, yao2022fundamental}. The relationship between speed and density is monotonically decreasing: as density increases, speed decreases.

This same observable is frequently employed for characterizing and designing pedestrian systems, which also exhibit the decreasing trend of mean speed with density \cite{kretz2019overview,aghamohammadi2018dynamic,bosina2018new, lohner2018fundamental,zhang2012pedestrian,jin2017large,johansson2009constant,helbing2007dynamics, seyfried2006basics,seyfried2005fundamental,weidmann1993transporttechnik,mori1987new, navin1969pedestrian,older1968movement,hankin1958passenger,ren2021flows, fruin1971pedestrian, predtechenskii1978planning, dinenno2008sfpe, hurley2015sfpe}.

While fundamental diagrams for vehicles and pedestrians are qualitatively similar, there is a significant distinction. For vehicles, contact is strictly prohibited. However, for pedestrians, contact between individuals naturally occurs at high densities. In particularly dense situations, pushing or shoving may be observed, especially in competitive behaviors \cite{lohner2018fundamental, helbing2007dynamics, garcimartin2017pedestrian}.

In vehicular dynamics, the decrease in speed can be attributed to the driver's voluntary speed reduction, either by applying the brakes or easing off the accelerator, to maintain a safe distance from the vehicle ahead. This mechanism holds across the entirety of the vehicular fundamental diagram. For pedestrians, a similar mechanism operates at low and medium densities: individuals reduce their desired speed as available space becomes limited. However, when physical contacts become prevalent due to increasing density, the nature of speed reduction shifts, rooted in physical constraints.

Thus, we posit that there are at least two mechanisms shaping the speed-density relation in pedestrian dynamics. One operates at low to medium densities, where remote sensing (like vision) influences the desired speed, which, in turn, affects the self-propulsion intensity, akin to vehicular traffic. The other, at high densities, involves short-range limitations imposed by physical constraints that curtail movement. The distinction between these two mechanisms has not been previously emphasized in the literature.

To address this oversight, we aim to experimentally study a straightforward system of physical agents that interact exclusively through contact, without any capacity for remote sensing. This approach will help us understand and isolate the section of the fundamental diagram related to contacts and physical space limitations, at least within this minimal system. We used commercially available vibration-driven vehicles (VDV) arranged in a one-dimensional setup (a circular track).

Furthermore, another objective of this research is to establish a precedent for utilizing the fundamental diagram as an insightful metric for a broad spectrum of active matter systems. This applicability goes beyond the confines of transport engineering, covering a spectrum of entities from animals and cells to robots and VDVs. In fact, the particular VDVs we utilized have already been employed to explore diverse facets of emergent behavior \cite{patterson2017clogging, patterson2020properties, barois2019characterization, dauchot2019dynamics, yang2020robust, binysh2022active}.

\section{Experimental setup}
\label{sec:Exp}

\subsection{VDVs and circular track}

\begin{figure*}
    \centering
    \includegraphics[width=\textwidth]{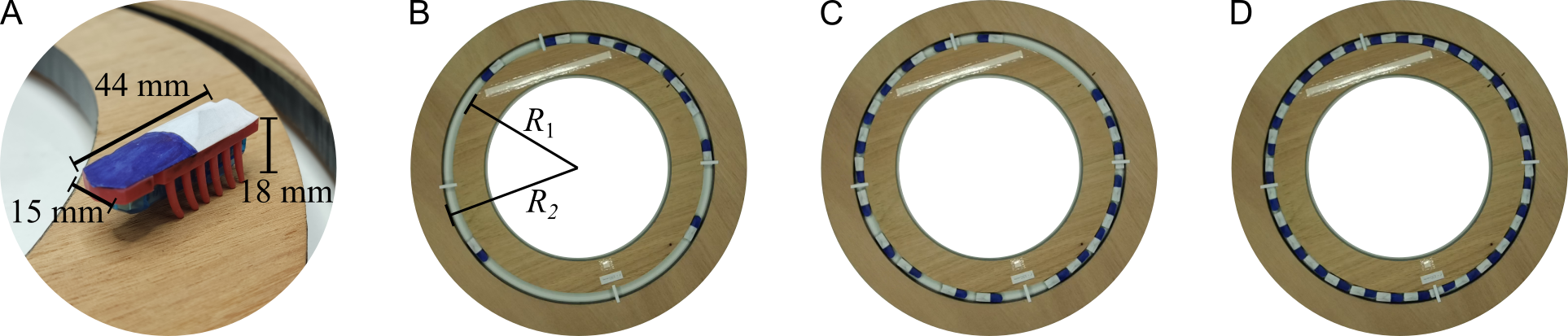}
    \caption{Experimental setup: A) VDV photograph and dimensions. A blue-painted label enables the tracking of trajectories. B) Snapshot of the system with $N=10$ VDVs. Two wooden rings are used to construct the circular track. C) Snapshot of the system with $N=20$ VDVs, and D) $N=30$.}
    \label{fig:setup}
\end{figure*}

The VDVs used in our experiments are the Hexbug Nano \cite{hexbug} (Fig.~\ref{fig:setup}A). Their body, with dimensions of $44 \times 15 \times 18\ \mathrm{mm}$, houses a vibrating motor inside. These vehicles stand on an asymmetrical bristle that rectifies the motor's vibration, resulting in forward movement. The motor is powered by a $1.5\ \mathrm{V}$, which, under normal usage conditions, provides over 90 minutes of operation. In the absence of a container or obstacle, the VDVs exhibit diffusive behavior in terms of their orientation of movement. In this study, we restricted their movement to a one-dimensional track by confining them within a circular channel with a width of $18\ \mathrm{mm}$. To achieve this, we used two concentric circular rings: the first with an outer radius $R_1 = 200\ \mathrm{mm}$, and the second with an inner radius $R_2 = 218\ \mathrm{mm}$, as shown in Fig.~\ref{fig:setup}B. The wooden ring walls were covered with a PVC sheet to ensure a uniform contact surface with the VDVs. The effective length of the channel, approximately $1313\ \mathrm{mm}$, allowed us to introduce a maximum of 30 VDVs, as can be seen in Figs.~\ref{fig:setup}B-D.

The experiments were recorded at a rate of 24 fps with a zenith camera. We used a white label, with the front painted blue, placed on each agent for tracking the VDVs (Fig.~\ref{fig:setup}A). By employing a standard blob detection method, we obtained the positions of the VDVs in each frame. Finally, velocities were calculated using a fourth-order finite difference method.

\subsection{Speed characterization}
\label{sec:IndSpeed}

First, we carried out an individual speed characterization for each of the 37 available VDVs. For this purpose, we conducted 180-second experiments for each VDV using fresh battery. We measured the average speed of each one and discarded the fastest and slowest ones. In this way, we ended up with a set of 30 VDVs whose speeds were uniformly distributed in the range of $90-120\ \mathrm{mm/s}$.

\section{Results}
\label{sec:Res}

To investigate the velocity response as a function of density, experiments were conducted over a duration of $180$ s for varying numbers $N$ of VDVs. The trials spanned densities with $N \in \{5, 10, 15, 20, 25, 30\}$. Specifically, the initial run started with $N=5$ VDVs. After $180$ s, another $5$ VDVs were introduced, and this increment continued until there were $N=30$ VDVs active during the concluding $180$ s interval. Thus, the total duration for this series of experiments, spanning these densities, was $T = 6\times 180\ \mathrm{s} = 1080\ \mathrm{s}$.

Three distinct sorting configurations were also examined:

1 - In the 'ascending' configuration, the initial $N=5$ VDVs were those with the lowest individual speeds, as detailed in Sec. \ref{sec:IndSpeed}. Subsequent introductions added the next fastest set of $5$ VDVs until $N=30$ was attained.

2 - Conversely, in the 'descending' configuration, the process started with the fastest $N=5$ VDVs. The next slower group of $5$ VDVs was then introduced, and so on.

3 - A third configuration involved randomly selecting VDVs without considering their individual velocities, to study the effects of increasing densities.

\subsection{Global Velocities}

\begin{figure}
    \centering
    \includegraphics[width=.45\linewidth]{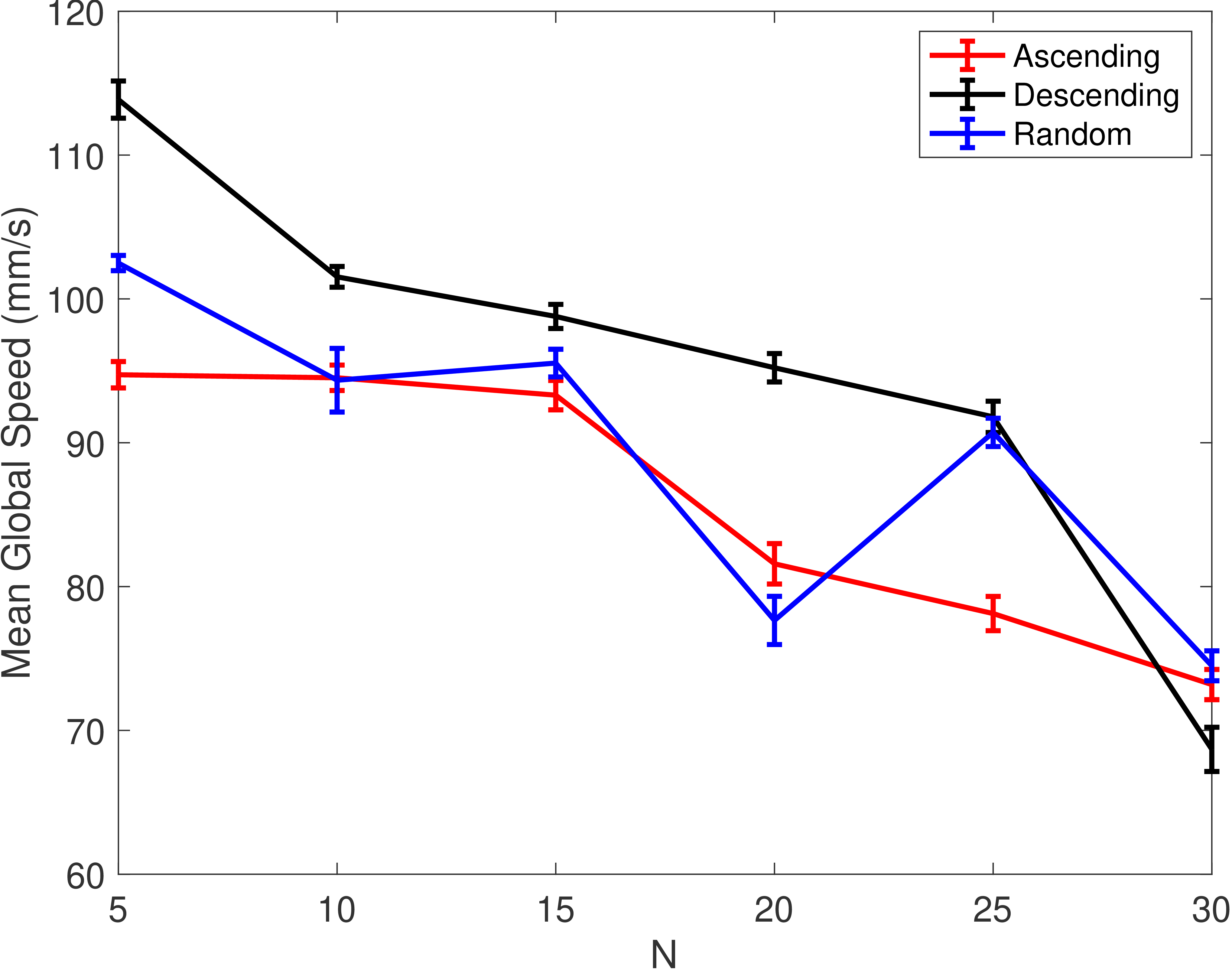}
    \caption{Mean speed as a function of $N$, for three sorting configurations. }
    \label{fig:VelG}
\end{figure}

As our first observable, we will calculate the average global velocities of the VDVs as a function of $N$. The global velocity is defined as $v^g_i = \frac{L_i}{180~ \text{s}}$, where $L_i$ is the distance traveled by particle $i$ during the $180$ s of the experiment. Figure \ref{fig:VelG} displays these quantities across the three scenarios: 'ascending', 'descending', and 'random'.

As expected, for small values of $N$, the 'ascending' scenario (where slower VDVs are introduced first) exhibits a lower global velocity compared to the 'descending' scenario (where the fastest VDVs are introduced first). In both cases, a decreasing trend in global velocity is observed as the number of VDVs increases. Furthermore, at maximum density ($N=30$), all scenarios converge to a similar average global velocity. This convergence suggests that, when reaching saturation and all VDVs are present, the order of introducing particles based on their speeds no longer seems relevant.

The 'random' scenario, in which VDVs are introduced without any specific order, appears to fall between the two extreme scenarios of 'ascending' and 'descending'. This observation supports the idea that random configurations are bounded by these two ordered configurations.

In the individual characterization, as discussed in Sec.\ref{sec:IndSpeed}, we found that the slowest vehicle had a speed of 90 mm/s. It is noteworthy that, with an increasing number of vehicles on the track, the overall speed decreases to values lower than that of the slowest VDV. This illustrates that the system's speed is not constrained by the slowest vehicle but rather exhibits a collapsing behavior

For a deeper understanding, the subsequent sub-sections will delve into the microscopic states of the particles.

\subsection{Distribution of Densities}
\label{sec:DensDist}

\begin{figure}
    \centering
    \includegraphics[width=1\linewidth]{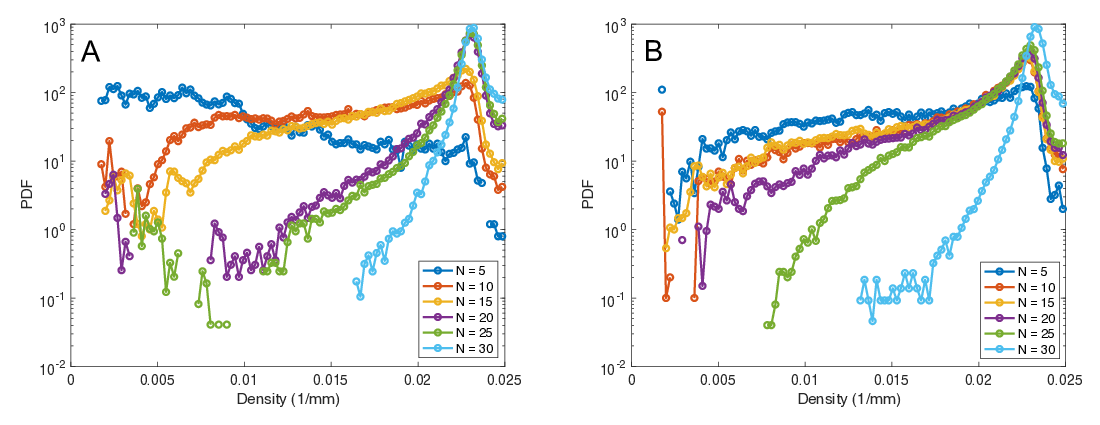}
    \caption{Distribution of individual densities. A) velocity 'ascending' order. B) velocity 'descending'. }
    \label{fig:FigDens}
\end{figure}

We compute the individual density of each particle $i$ using the relation $\rho_i= 1/d_{ij}$, where $d_{ij}$ represents the distance along the circular path to the nearest neighboring particle. In Fig. \ref{fig:FigDens}, the probability density function (PDF) of densities, for all particles and all times, is displayed for each experiment with varying numbers of VDVs, encompassing both the 'ascending' and 'descending' scenarios.

For all cases, it can be observed that there is a peak in the distribution around $\rho=0.023$ 1/mm. This value aligns with the minimum theoretical distance between two contacting VDVs, given that each has a maximum length of $44$ mm. However, tails of the distributions can be seen extending toward higher density values. This can be attributed to the triangular geometry of the VDV's front and back. Slight misalignments, possibly due to the curvature of the track, might allow for distances shorter than the theoretical minimum. Other factors contributing to these increased densities could include overlapping of VDVs (one mounting over another) and fluctuations in the automatic tracking system's detection of the label.

Another shared characteristic between the 'ascending' and 'descending' scenarios is that the probability in the low-density zone of the distribution diminishes as $N$ increases.

In the peak zone, the 'ascending' scenario (Fig. \ref{fig:FigDens}A) shows a clear separation between the three greater values of $N={20, 25, 30}$ and the smaller ones $N={5,10, 15}$. The former set tends to exhibit a heightened probability aligned with the maximum theoretical density. In the 'descending' scenario (Fig. \ref{fig:FigDens}B), this distinction is apparent only for $N=30$. 

Overall, as the values of $N$ increase, the distributions tend to become more pronounced, peaking at the maximum theoretical density dictated by the VDV's geometry.

\subsection{Distribution of Velocities}
\begin{figure}
    \centering
    \includegraphics[width=1\linewidth]{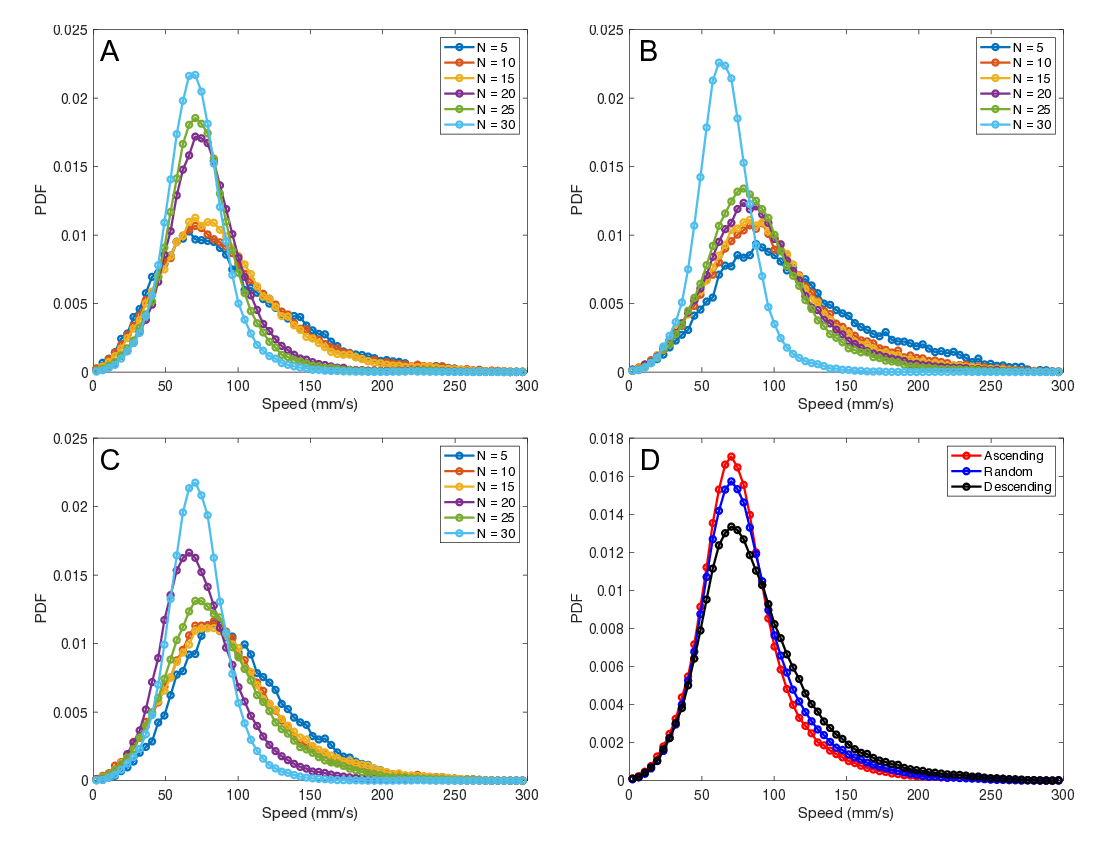}
    \caption{Distribution of individual velocities. A) VDVs introduced in an 'ascending' sequence according to their free velocity. B) The same, but in a 'descending' sequence. C) VDVs added without any specific order. D) PDF combining all experiments with varying $N$, grouped solely by the type of order of inclusion.}
    \label{fig:FigVels}
\end{figure}

Complementarily, we examine in detail the distribution of microscopic velocities for each VDV and every time frame in Fig. \ref{fig:FigVels}.

Similar to the analysis of densities, the 'ascending' scenario depicted in Fig. \ref{fig:FigVels}A shows a distinction between the velocity distributions for larger ($N={20, 25, 30}$) and smaller ($N={5,10, 15}$) number of VDVs. This observation suggests a marked shift toward lower velocities when there are more than 15 VDVs in the system. In the 'descending' order scenario (Fig. \ref{fig:FigVels}B), only the $N=30$ case distinctly separates from the others, indicating that when VDVs are introduced from fastest to slowest, the system's velocity substantially decreases only at a very high density ($N=30$). Moreover, when comparing the 'ascending' and 'descending' configurations, one can observe that the peak positions in the distributions shift toward higher speed values, consistent with the average global velocity presented in Fig. \ref{fig:VelG}.
The random configuration, as illustrated in Fig. \ref{fig:FigVels}C, does not exhibit a distinct separation in the PDF, positioning itself between the two extreme ordered configurations. This becomes more evident in Fig. \ref{fig:FigVels}D, which considers the distribution across all $N$ values. In the 'ascending' scenario, there is a higher probability for low velocities and a reduced likelihood for high velocities. Conversely, in the 'descending' configuration, there is a heightened probability for low velocities and a diminished chance for higher velocities. Through this analysis, it is evident that the random order configuration gives a PDF that falls between the two extremes.

Another noteworthy aspect is the distribution width. It can be observed that in experiments with higher values of $N$ (Figs.~\ref{fig:FigVels})A-C, speed distributions become narrower. This phenomenon occurs because, as the number of vehicles in the channel increases, each particle experiences limited movement, resulting in a more uniform rotation speed.

\subsection{VDVs Fundamental Diagram}

\begin{figure}
    \centering
    \includegraphics[width=1\linewidth]{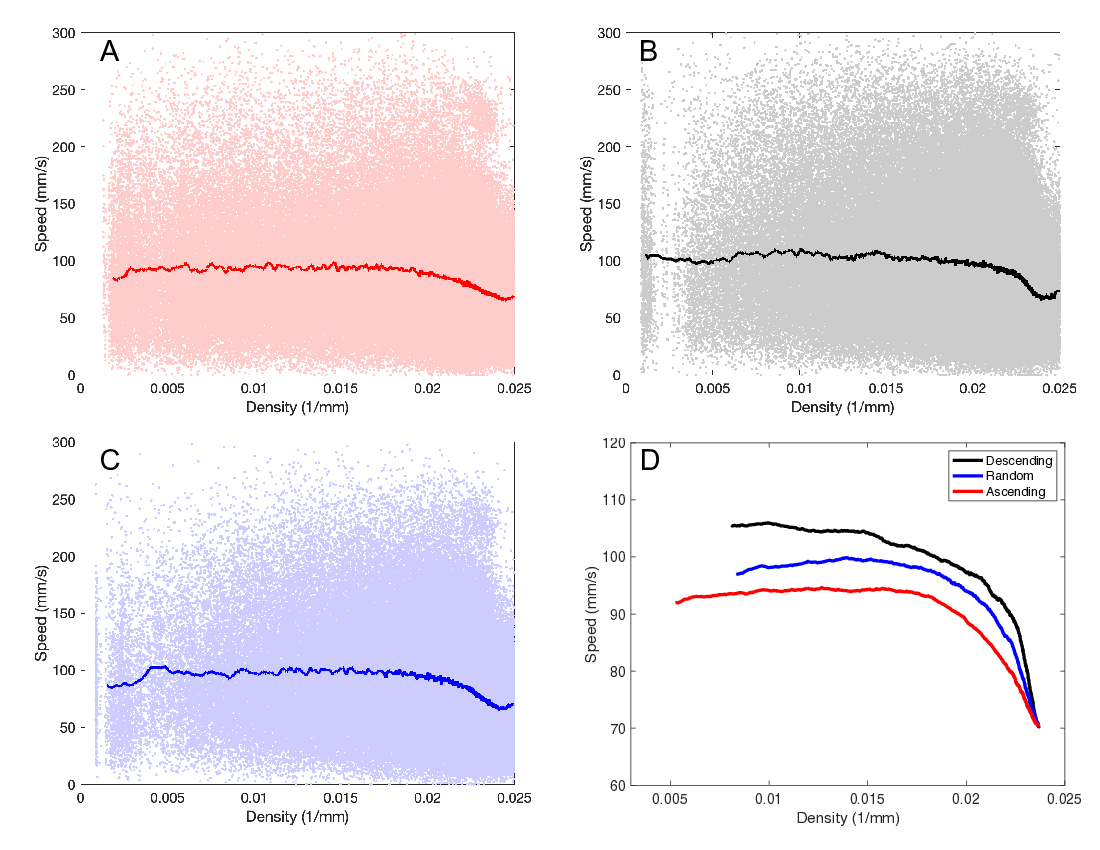}
    \caption{Microscopic speed-density data. A) Vehicles introduced in an 'ascending' order. B) 'Descending' order. C) 'Random' order. D) Moving average of the three aforementioned data sets, calculated over 20,000 data points.}
    \label{fig:Fig_FD}
\end{figure}

Now, after characterizing microscopic densities and velocities separately, we delve into the speed-density diagram, as presented in Fig.~\ref{fig:Fig_FD}. The first three panels, Figs.~\ref{fig:Fig_FD}A, B, and C, display the entire collection of microscopic data, encompassing all values of $N$ and all time instances (with a total of $F_n = 1080 \times 24 = 25920$ analyzed frames). As is clear, there is a wide dispersion of data points, lacking a clear trend. However, when a moving average of 1000 data points is plotted, a subtle decrease in speed becomes discernible at higher density values. Still, overall, these averages remain relatively flat. At the highest densities, a minimum, which corresponds to the theoretical maximum density determined by the VDV's geometry, is observed in all three cases. The upward curve to the right of these minima can be attributed to, misalignment, overlapping and tracking errors, as explained in Sec.~\ref{sec:DensDist} in reference to Fig.~\ref{fig:FigDens}.

By limiting the analysis to data up to the maximum theoretical density of $\rho = 0.023$ 1/mm and using a 20000-point moving average, we present the speed-density plot for the three configurations in Fig.~\ref{fig:Fig_FD}D. Three main observations can be drawn from this analysis.

First, as expected, the results are in accordance with those regarding the global velocity presented in Fig.\ref{fig:VelG}. In this context, the 'descending' configuration, which involves introducing the fastest VDV first, exhibits higher velocities when compared to the 'ascending' configuration at lower and moderate densities. However, as the density increases, both configurations gradually converge to the same velocity value at the maximum achievable density. Furthermore, the structure observed during the velocity PDF analysis for all values of $N$ are confirmed: the random order scenario distinctly lies between the two extreme ordered configurations, 'ascending' and 'descending'.

Secondly, the velocity profile remains relatively stable for low to medium densities until it approaches the density zone influenced by the geometric constraints of the vehicles. The significant inflection point occurs at approximately $\rho \sim 0.02$ 1/mm, which corresponds to an average distance of about 50 mm, roughly 11\% greater than the vehicle size. Since this value represents an average, contact events start from this typical distance and become more frequent as the mean distance between neighboring vehicles decreases. This clarifies how speed-density curves behave without long-range sensing, with speed reduction mainly due to agent collisions.

Finally, considering the reduction in speed looking at the speed-density curves (Fig. \ref{fig:Fig_FD}) along with the global speed  as a function of $N$ (Fig. \ref{fig:VelG}), we can establish that the maximum reduction in the mean speed ranges from 25\% to 40\%, depending on the arrangement of the vehicles.

\section{Conclusions}

In summary, we experimentally studied a system of self-propelled particles whose trajectories were, using a circular track, restricted to circular motion. These particles, due to small differences in their construction, have differences in the speed they can achieve. We found that as the number of VDVs increased, the average speed of the system decreased regardless of the order in which the vehicles were added. Moreover, the results showed that the average speed of the system, when considering the highest number of VDVs, is less than the lowest of the speeds of all vehicles.

We then analyzed the microscopic characteristics by means of the density and speed experienced by each VDV. The density distributions showed a maximum value that corresponds to the contact distance of two VDVs. In addition, as $N$ increases, the distributions become more pronounced and the population in the low-density region decreases significantly. Similarly, as the value of $N$ increased, the velocity distributions became narrower and more pronounced. This occurred because the reduced available space resulted in a more uniform system speed.

Compiling the microscopic data, we constructed the speed-density diagram for VDVs. Given that these agents lack long-range sensing and interact solely through contact, speed remains constant until the density reaches a point roughly 10\% beyond one agent's length, at which stage it significantly declines. This observation may provide insights into the behavior of other systems in situations where long-range sensing is absent, and speed reduction primarily arises from direct collisions between agents.

In a one-dimensional system, such as the one examined in this study, the behavior in the low and medium density ranges is influenced by the order in which agents are added to increase density. When fast VDVs are introduced first, they result in higher speed values on the fundamental diagram compared to the scenario where slower VDVs are added first. However, as the density increases, both curves gradually converge to a common speed value at maximum density. Additionally, we observed that both the speed-density curve and velocity distributions for any arbitrary order configuration fall between these two extreme ordered configurations.

From this observation, we can derive at least two significant insights:
(a) If this behavior remains consistent across all random configurations, then organizing the particles based on their free speed could provide an efficient method for capturing the full spectrum of potential speed-density curves, rather than repetitively conducting the same experiment with randomly ordered configurations. 
(b) Given the variability of speeds, the arrangement order of agents significantly influences the specific shape of the fundamental diagram, at least in one dimension. This factor may have been previously overlooked in attempts to explain the variation observed in different pedestrian fundamental diagrams.

\section*{ACKNOWLEDGMENTS}
This work was funded by Research Grant from HFSP (Ref.-No: RGP0053/2020) and project PICT 2019-00511 (Agencia Nacional de Promoci\'on Cient\'ifica y Tecnol\'ogica, Argentina).

\bibliographystyle{unsrt}  
\bibliography{references}

\end{document}